\documentclass[letterpaper,aps,pra,nolongbibliography,twocolumn,showpacs,floatfix,superscriptaddress,10pt]{revtex4-2} \usepackage[utf8]{inputenc}
\usepackage[english]{babel}
\usepackage{definitions}

\renewcommand{\tr}{\operatorname{tr}}

\usepackage[draft,inline,nomargin]{fixme} \fxsetup{theme=color}
\FXRegisterAuthor{cp}{acp}{\color{blue}CP}
\FXRegisterAuthor{ac}{acg}{\color{orange}AC}
\FXRegisterAuthor{en}{esn}{\color{red}EN}
\FXRegisterAuthor{dd}{ddg}{\color{purple}DD}
\usepackage{graphicx}\usepackage{dcolumn}\usepackage{bm}\newcommand{\sref}[1]{Sec.~\ref{#1}}
\newcommand{\eref}[1]{Eq.~(\ref{#1})}
\newcommand{\fref}[1]{Fig.~\ref{#1}}
\graphicspath{{"figures/"}}
\definecolor{refkey}{rgb}{0,0,1}
\definecolor{labelkey}{rgb}{0,1,0}
\begin{document}

\title{
Coarse-grained dynamics in quantum many-body systems using the maximum entropy principle
}\newcommand{\sas}{Institute of Physics, Slovak Academy of Sciences, D\'ubravsk\'a Cesta 9, Bratislava 84511, Slovakia}
\newcommand{\cucei}{Departamento de F\'isica, Universidad de Guadalajara, Guadalajara, Jalisco 44430, M\'exico}
\author{Adán Castillo}
\affiliation{\UNISTRA}\author{Carlos Pineda}\email{carlospgmat03@gmail.com}\affiliation{\IFUNAM}
\author{Erick Sebastián Navarrete}
\affiliation{\FCUNAM}\affiliation{\IFUNAM}\author{David Davalos}
\affiliation{\sas}
\affiliation{\cucei}

\begin{abstract}
Starting from a coarse-grained map of a quantum many-body system, we construct
the inverse map that assigns a microscopic state to a coarse-grained state
based on the maximum entropy principle. Assuming unitary evolution in the
microscopic system, we examine the resulting dynamics in the coarse-grained system using
the assignment map. We investigate both a two-qubit system, with \swap\ and
controlled-\textsc{not}\ gates, and $n$-qubit systems, configured either in an Ising spin
chain or with all-to-all interactions. We demonstrate that these dynamics
exhibit atypical quantum behavior, such as nonlinearity and non-Markovianity.
Furthermore, we find that these dynamics depend on the initial coarse-grained
state and establish conditions for general microscopic dynamics under which
linearity is preserved. As the effective dynamics induced by our coarse-grained
description of many-body quantum systems diverge from conventional quantum
behavior, we anticipate that this approach could aid in describing the
quantum-to-classical transition and provide deeper insights into the effects of
coarse graining on quantum systems.
\end{abstract}
\maketitle
\section{Introduction} Coarse-grained descriptions have been extensively used to
study quantum systems~\cite{PhysRevLett.99.180403, PhysRevLett.107.250401, PhysRevA.85.042115, PhysRevA.101.032303, PhysRevLett.127.120401, e19050207, PhysRevA.84.022103, PhysRevResearch.5.043057, PhysRevA.102.032217, 
PhysRevA.100.042114}. 
This type of effective modeling is essential because the full quantum description is
usually impractical as the Hilbert space dimension grows exponentially with respect
to the number of particles.
In this regard, coarse graining is generally implemented by reducing the
number of degrees of freedom; however, the way this is done is not
unique~\cite{Ziman2008}.
Moreover, coarse graining has been pivotal in studying the quantum-to-classical
transition \cite{Busch1993,PhysRevLett.99.180403}, which remains, along with the
measurement problem, an active area of research due to the inherent linearity
of the theory and the mathematical connections between the studied entities and
the outcomes of measurement~\cite{Zur91,RevModPhys.76.1267,SCHLOSSHAUER20191}.
It has been employed also in the study of the thermalization of closed quantum
systems~\cite{PhysRevA.99.010101,PhysRevA.99.012103}. \par 
Using the language of quantum information, several authors have introduced the
concept of  coarse graining in the quantum
realm. In Ref.~\cite{Ziman2008} coarse graining was defined via classical
stochastic maps between measures defined using quantum observables; in
Ref.~\cite{CGEmergingDynamics} a framework exploiting Stinespring
dilation~\cite{Stinespring2006} is introduced to keep track of the information
dumped. Both frameworks accommodate a variety of coarse-graining maps. A 
class of quantum channels, amounting to a coarse-grained description 
of many-body systems, was introduced in \cite{FuzzyMeasurements}. A key
feature in common is that the coarse-grained description contains less
information than the original fine-grained picture. 
\par

A natural and fundamental question concerns how a physical system appears when
probed with imperfect measurement devices. This issue is of central importance,
as all experimental observations are inherently limited by some degree of
imprecision. Such limitations raise critical concerns about the implications
they may have on the information extracted from a system. In special
relativity, for example, the behavior of rods and clocks plays a key role in
defining foundational concepts such as simultaneity~\cite{Einstein1905}, while in
quantum mechanics, the von Neumann postulate establishes the framework for what
can be observed and how~\cite{vonNeumann1955}. In the context of quantum states, it has been
shown that the space of observable states shrinks doubly exponentially as a
result of finite addressing errors~\cite{FuzzyMeasurements}. Of particular interest, both
fundamentally and practically, is the question of how the dynamics of many-body
quantum systems appears through the lens of imperfect measurements. The
limitations introduced by such coarse-grained observations must be carefully
considered when analyzing and interpreting the evolution of complex quantum
systems.

This question has been addressed from several points of
view.  There have been efforts to define and investigate the dynamics governing
coarse-grained descriptions of quantum systems. 
For example, in
Ref.~\cite{CGEmergingDynamics} the impact on the emerging dynamics due to the
choice of the fine-grained state was discussed. In Refs.~\cite{Macro-To-Micro,
Vallejos_2022} two ways of assigning fine-grained quantum states were analyzed
for the blurred detector across various assignment maps. The key result in
these works is the emergence of non linear and non-Markovian dynamics. We
believe that a deeper exploration of such results is needed, especially in the
context of many-body systems under general but tractable enough assumptions. \par
In this work we address this need by exploring coarse-grained descriptions of
many-body quantum systems and the emergent dynamics using the idea of
assignment map~\cite{Macro-To-Micro, Vallejos_2022} to ``invert'' the
coarse-graining map.  Two crucial objects must be chosen carefully to achieve
our goal. First, we need a general enough coarse-graining map to accommodate
a decent variety of many-body systems. 
To do this, we consider a recently
developed coarse-grained model~\cite{FuzzyMeasurements}, which is based on
particle-indexing noise and a reduction of the number of particles.
More specifically, this model assumes that when a measurement is
performed, possibly involving multiple particles, the subset of particles being
measured can vary randomly. This coarse-graining approach has two key features
that make it relevant for further study. First, it captures the limitations of
a measurement device that cannot resolve individual particles, resembling the
behavior of a macroscopic apparatus observing a quantum system. Potentially,
this can contribute to a better understanding of the quantum-to-classical transition.
Second, the model allows for the exploration of how different spatial
configurations influence quantum states. It can incorporate various geometries,
such as linear chains (see, e.g.,~\cite{PhysRevResearch.6.023096,Cho2024}),
two-dimensional lattices (see, e.g.,~\cite{Yu2024,Davis2023}), or more complex layouts like
those found in IBM quantum processors (see, e.g.,~\cite{PRXQuantum.5.040334}).

Second, we must choose an assignment map. There is a variety of criteria for
selecting one. In Ref.~\cite{Vallejos_2022} the authors analyzed the difference
between two assignment maps, one on them based on the maximum entropy
(MaxEnt) principle and the other based on the average compatible pure states.  The MaxEnt
principle serves as a method of inference that assigns probabilities to all
events while ensuring compatibility with the known information, given as
expectation values.  The resulting distribution is the least biased, i.e., it
maximizes the entropy~\cite{shannon}.  The role of the MaxEnt principle as an
inference method in the formulation of statistical mechanics ensembles for
general observables was thoroughly discussed in Jaynes' widely cited
paper~\cite{PhysRev.106.620}. This work extends beyond the canonical and
microcanonical ensembles, providing a more general framework.  Its application
to the quantum realm was further developed in a subsequent
paper~\cite{PhysRev.108.171} (see also~\cite{10.1063/1.1704014}).  In the
quantum case, the measure of uncertainty is the von Neumann entropy and the
resulting probability distribution (or density matrix, in the quantum case)
reflects these values without introducing any additional information.  It is
worth mentioning that this principle has been used for tackling the lack of
information in other scenarios, for example, in incomplete state (and process)
quantum tomography~\cite{MarioMaxEnt, BusekMaxEnt, buzekspins}.  We have chosen
MaxEnt to construct the assignment map compatible with the coarse-graining map
for both its solid physical motivation, in terms of information, and the fact
that it is analytically convenient. 

Having established the connection between both scales using the aforementioned
maps, we investigate the emergent coarse-grained dynamics induced by various
quantum dynamics at the finer scale. Specifically, we assume that only the
coarse-grained scale is accessible, meaning that expectation values are
expressed in terms of operators at this scale. Consequently, we construct the
corresponding operators at the finer scale such that they reproduce the same
expectation values while simultaneously encoding the coarse-graining map. With
these elements in place, we formulate the MaxEnt states required for the
assignment map. Finally, the emergent coarse-grained dynamics naturally arise
as the concatenation of the coarse-graining maps and the dynamics at the finer
scale.
We consider examples of both  small and large many-body quantum systems.

The paper is organized as follows. In Sec.
\ref{sec:CGMaxEntEffDyn} we introduce both the coarse-graining map and the
maximum entropy assignment map, and with these elements we proceed to construct the
effective dynamics of the macroscopic system. Then we apply these tools to
different quantum dynamical systems.  In particular, Sec.
\ref{sec:EffectiveGates} is devoted to the study of the effective dynamics of a
two-qubit system with underlying \swap\ and controlled-\textsc{not} gates. In Sec.
\ref{sec:SpinChains} we explore the effective dynamics of an $n$-qubit system
configured as an Ising spin chain, with both all-to-all and nearest-neighbor
interactions. In Sec. \ref{sec:LinearDynamics} we establish
conditions for general microscopic dynamics under which linearity is preserved,
and we present an example of a microscopic dynamics that preserves linearity
but is nonetheless non-Markovian. We summarize our work and discuss our conclusions
in~\sref{sec:conclusions}.\par
\section{Effective dynamics in coarse-grained systems} \label{sec:CGMaxEntEffDyn} In this section we introduce the coarse-graining map from which the effective
description of the state is recovered. We construct the maximum entropy
assignment map that will be used to assign a microscopic state to the effective
state, and study some of its properties. This map is used to formally present
the effective dynamics in a quantum many-body system subject to a coarse
grained description. 
\par
\subsection{The coarse-graining map} 

Consider an imperfect measuring device that is subject to two types of errors.              
First, the device lacks the capability to accurately distinguish between the
individual particles of a $d$-level $n$-body system. In other words, there
exists a non-zero probability  $p_{P}$ of wrongly identifying particles
according to a permutation $P$. This type of measurement is referred to as a
fuzzy measurement.
Second, the device does not have the ability to resolve all particles; only
a subset $\tau$ of $m$ particles can be measured.
Mathematically, a partial trace over the complement of $\tau$ is performed.
A coarse-grained description of the system is obtained when both
types of error are combined~\cite{FuzzyMeasurements}, in the sense
that expected values should be calculated with respect
to an effective state resulting from the application of the
coarse-graining map
\begin{equation}\label{eq:GeneralCG}
    \begin{gathered}
        \mcC:\mcB( \hilbert_{d}^{\otimes n})\rightarrow\mcB( \hilbert_{d}^{\otimes m}),\\
        \varrho \mapsto \tr_{\overline{\tau}}\qty(\sum_{P}p_{P}P(\varrho)),
    \end{gathered}
\end{equation}
where $\tr_{\overline{\tau}}$ represents the partial trace over the complement
of $\tau$ and $\mcB(\hilbert)$ is the space of bounded linear operators acting on $\hilbert$.\par

Our primary focus is directed towards a particular case of the
coarse-graining map given by \eqref{eq:GeneralCG}. First, the measuring device
might mistakenly swap pairs of particles. Second, it is able to resolve a single
particle, that is, $m=1$. Without loss of generality,
we can assume that the first particle is the one that is intended
to be measured, so that $\tau=1$. 
The coarse-graining map that captures this situation is
\begin{equation}\label{eq:CG}
    \begin{gathered}
        \mcC:\mcB( \hilbert_{d}^{\otimes n})\rightarrow\mcB( \hilbert_{d}),\\
        \varrho \mapsto \tr_{\overline{1}}\qty(\sum_{k}p_{k}P_{1,k}(\varrho)),
    \end{gathered}
\end{equation}
where $P_{1,k}$ is the permutation between the first and the $k$th particle and
$P_{1,1}\equiv\Id$. If state tomography were to
be performed on the system using our imperfect apparatus, the
result would be a one-particle effective state
$\rho_{\text{eff}}=\mathcal{C}(\varrho)$; similarly, the
expected value of any single-particle observable $A$ will be $\tr{[A
        \mathcal{C}(\varrho)]}$. It is crucial to recognize that the resulting effective
state is a mixture of all the reduced systems of the
microscopic system (assuming $p_{k}\neq 0\, \forall\, k$).\par

Two specific probability distributions $p_{k}$ hold particular significance in
this context:
\begin{align}\label{eq:ProbDistributions}
    p_{k}=\frac{1}{n}\,\forall\,k, \quad p_{k}=\frac{1-p_{1}}{1-n} \,\forall\,k\neq 1.
\end{align}
The first one represents the scenario in which the measuring device is
completely incapable of distinguishing between the individual particles. This
situation is akin to a gas, where each particle has an equal probability of
being found anywhere within the system. The second one can be interpreted as
the first particle being the one that is closest to the measuring device and
the other providing a mean background noise. These
distributions are referred to as non preferential and preferential
distributions, respectively.\par

\subsection{Maximum entropy principle and the assignment map}

Now let us shift our attention to the assignment map, which allows us to make
an inference about the microscopic state of the system through our
coarse-grained measurements. It is through this map that we will be able to
obtain the state that will evolve according to the microscopic dynamics.  The
assignment map is constructed by selecting the microscopic state that maximizes
the von Neumann entropy, while ensuring that it corresponds to the effective
macroscopic state under a specific coarse-graining map, in a similar spirit
as in~\cite{PhysRev.106.620,dysonCircular} or in the context of coarse
graining~\cite{CGEmergingDynamics, Vallejos_2022}. The maximization of
entropy is carried out using the method of Lagrange multipliers, where the
information of the effective state will be used as constraints. \par

Let us assume that we are able to perform state tomography on the effective
state~\cite{Chuang}.
If $\{\varsigma^{\alpha}\}_{\alpha}$ is a tomographically complete set of 
operators acting on the Hilbert space $\hilbert_{d}$ of the effective state 
$\mcC(\varrho)$, then its expected values can be connected to 
the expected values of a set of operators $\{G^{\alpha}\}_{\alpha}$ 
acting on the Hilbert space of the microscopic state, 
$\hilbert_{d}^{\otimes n}$. These operators are defined by 
\begin{equation}\label{eq:GOperators}
    G^{\alpha} = \sum_{k=1}^{n}p_{k}\varsigma_{k}^{\alpha},
\end{equation}
where $\varsigma^{\alpha}_k$ is the operator $ \varsigma^{\alpha}$
acting on the $k$th particle. 
This connection is expressed mathematically by
\begin{equation}\label{eq:PauliG}
    \tr(\varsigma^{\alpha} \rho_{\eff}) =\tr(G^{\alpha}\varrho)
\end{equation}
and is studied in Appendix \ref{ap:FuzzyOperators}.
Notice that the operators $G^{\alpha}$ can be thought of as 
fuzzy operators since they are equivalent to applying
the $\varsigma^{\alpha}$ operator to the $k$th particle with probability $p_{k}$.
Note that the set $\{G^{\alpha}\}_\alpha$ 
is not tomographically complete, implying that
we do not have access to all the details about the microscopic system. \par

Given the expected values of the $G^{\alpha}$ operators, the state that maximizes
the von Neumann entropy is then~\cite{10.1063/1.1704014}
\begin{equation}\label{eq:UnsimplifiedMaxEntState}
    \varrho_{\max}=\frac{1}{Z}\text{exp}\qty(\sum_{k=1}^{n}p_{k}\sum_{\alpha=1}^{d^{2}-1}\lambda_{\alpha}\varsigma_{k}^{\alpha}),
\end{equation}
where $\lambda_{\alpha}$ is the Lagrange multiplier associated with the expectation value
$\tr(\varsigma^{\alpha}\rho_{\ef})$. Indeed,
\begin{equation}\label{eq:ExpValLambda}
    \tr(\varsigma^{\alpha}\rho_{\eff})=\frac{\partial}{\partial \lambda_{\alpha}}\ln(Z),
\end{equation}
with the partition function
\begin{equation}
    Z=\tr\qty[\text{exp}\qty(\sum_{\alpha=1}^{d^{2}-1}\lambda_{\alpha}G^{\alpha})].
\end{equation}
Because the operators
$p_k\sum_\alpha\lambda_{\alpha}\varsigma_{k}^{\alpha}$ in Eq. \eqref{eq:UnsimplifiedMaxEntState}
commute, it is easy to observe that the maximum entropy state is separable.
Due to this property, the maximum entropy assignment map is 
\begin{equation}\label{eq:MaxEntAss}
    \begin{gathered}
        \mcA_{\mcC}^{\max}:\mcS( \hilbert_{d})\rightarrow\mcS( \hilbert_{d}^{\otimes n}),\\
        \rho_{\ef} \mapsto \Motimes_{k=1}^{n}\frac{1}{Z_{k}}\text{exp}\qty(p_{k}\sum_{\alpha=1}^{d^{2}-1}\lambda_{\alpha}\varsigma^{\alpha}),
    \end{gathered}
\end{equation}
where $Z_{k}$ is the partition function associated with the 
$k$th particle and $\mcS( \hilbert)$ is the set of density matrices acting on 
$\hilbert$. The dependence of the assigned state on the effective 
state is encoded in the $\lambda_{\alpha}$ parameters. 
In fact, given the effective state $\rho_{\ef}$, the 
$\lambda_{\alpha}$ parameters are obtained by noting that the norm of the 
generalized Bloch vector of the effective state, $\vec{r}_\ef=r_{\ef}\hat{n}_{\ef}$, 
can be expressed as
\begin{equation}\label{eq:GeneralizedBlochVectorNorm}
    r_{\ef}=\sum_{k=1}^{n}p_{k}r_k,
\end{equation}
where
\begin{align}\label{eq:LambdaNorm}
    r_k=\tanh{(p_{k}\lambda)}, \quad \lambda = \qty(\sum_{\alpha}\lambda_{\alpha}^{2})^{1/2}.
\end{align}
Thus, inverting \eqref{eq:GeneralizedBlochVectorNorm}, we can find all the 
$\lambda_{\alpha}$ parameters from $\vec{r}_{\ef}$. More precisely,
\begin{equation}
    \lambda_{\alpha} = \frac{r_{\ef}^{\alpha}}{r_{\ef}}\lambda,
    \label{eq:lambdas}
\end{equation}
where $r_{\ef}^{\alpha}$ is the $\alpha$ component of the $\vec{r}_{\ef}$ Bloch vector.
Now the assigned states by the maximum entropy assignment map in the preferential and 
non preferential cases \eqref{eq:ProbDistributions} are
\begin{align}
    \mcA_{\mcC}^{\max}(\rho_{\ef})=&\rho_{1}\otimes\rho_{\text{np}}^{\otimes(n-1 )} \label{eq:PreferentialAss},\\
    \mcA_{\mcC}^{\max}(\rho_{\ef})=&\rho_{\ef}^{\otimes n},
    \label{eq:NonPreferentialAss}
\end{align}
respectively. In the preferential case \eqref{eq:PreferentialAss}, we have defined 
the reduced state of the non-preferential particles as
\begin{equation}\label{eq:NonPrefReduced}
    \rho_{\text{np}}=\frac{1}{Z_{\text{np} }}\exp\qty(\frac{1-p_{1}}{n-1}\sum_{\alpha=1}^{d^2-1}\lambda_{\alpha}\sigma^{\alpha}).
\end{equation}

Noticeably, the assignment map $\mcA_{\mcC}^{\max}$ is clearly non linear [see~\eref{eq:NonPreferentialAss}]. As an example, consider the non preferential case with the state $\Id/2$; we have $\mcA_{\mcC}^{\max}(\Id/2)=\Id/2\otimes \Id/2=\Id/4$, which is not equal to $\mcA_{\mcC}^{\max}(\dyad{0})/2+\mcA_{\mcC}^{\max}(\dyad{1})/2=\dyad{00}/2+\dyad{11}/2$.
Also observe that $\mcA_{\mcC}^{\max}$ is defined once $\mcC$ is given such
that their interplay is 
\begin{equation}
\mcC\circ\mcA_{\mcC}^{\max}=\text{id}_{\mcS(\hilbert_d)},
\label{eq:consistency}
\end{equation}
where the map $\text{id}_{\mcS(\hilbert_d)}$ is an affine mapping and has a
unique linear extension on $\mcB(\hilbert_d)$~\cite{Ziman2008}, namely,
$\text{id}_{\mcB(\hilbert_d)}$. This is relevant because, as we will show
later, the emergent effective dynamics is generally non linear.  The
non linearity will come from the interplay of $\mcC$ and $\mcA_{\mcC}^{\max}$,
with the microscopic dynamics, where the only element that is non linear is
$\mcA_{\mcC}^{\max}$ alone.

\subsection{The effective dynamics} 

Now that we have established the use of a coarse-graining model that
incorporates both resolution and permutation errors and have constructed
an assignment map based on the maximum entropy principle, we can explore
the evolution of the effective system.
Since we assume that the state that propagates due to the underlying evolution
is precisely the one assigned via the maximum entropy assignment map, we can 
represent the dynamics of the effective state through the commutative
diagram
\[\begin{tikzcd}[arrows={<-|}]
    \rho_{\ef}(0)  & \rho_{\ef}(t) \arrow{l}{\Gamma_{t}} \arrow{d}{\mcC}\\
    \varrho_{\max}(0) \arrow{u}{\mcA_{\mcC}^{\max}} & \varrho_{\max}(t), \arrow{l}{\mcV_{t}}
\end{tikzcd}
\]
where $\mcV_{t}$ is the evolution of the microscopic state. The evolution
of the effective state $\Gamma_t$ can be thus obtained from the composition
three different operations \cite{Macro-To-Micro}. First, a microscopic 
state is assigned to the effective state through the MaxEnt assignment map. 
Then the assigned map is propagated via $\mcV_{t}$. Finally, we define the 
coarse-grained description of the evolved assigned state to be the evolved 
effective state. In this way, we define the coarse-grained dynamics as the 
composition
\begin{equation}\label{eq:GammaMap}
    \begin{gathered}
        \Gamma_{t}: \mcS( \hilbert_{d})\rightarrow\mcS( \hilbert_{d}),\\
        \rho_{\ef} \mapsto (\mcC\circ\mcV_{t}\circ\mcA_{\mcC}^{\max})(\rho_{\ef}).
    \end{gathered}
\end{equation}

As mentioned above, while \( \mathcal{C} \circ \mathcal{A}_{\mathcal{C}}^{\max}
\) admits the identity map as a linear extension on \( \mathcal{B}(\hilbert_d)
\), this linearity is generally broken in the full composition \( \mathcal{C}
\circ \mathcal{V}_t \circ \mathcal{A}_{\mathcal{C}}^{\max} \), giving rise to
the generally non-linear dynamics defined above. Although all
maps contribute to the composition, the only inherently non-linear map is \(
\mathcal{A}_{\mathcal{C}}^{\max} \) itself.
\subsection{Complete positivity and trace-preserving of the effective dynamics} In this section we prove that $\Gamma_t$ is a non linear completely positive and trace-preserving (CPTP) map. To do this, we first show that both the assignment map and the effective dynamics are homogeneous of degree 1. This property implies trace preservation and will aid in the proof that both $\Gamma_t$ and $\mathcal{A}_{\mathcal{C}}^{\max}$ are completely positive.
\subsubsection{Homogeneity and trace preservation} As $\mcC$ and $\mcV_t$ are linear, and therefore homogeneous of degree 1, it suffices to prove the homogeneity of $\mcA_\mcC^{\max}$ to conclude that $\Gamma_t$ is also homogeneous.

To proceed, consider an operator $\Delta = k \rho$, where $\rho \in \mathcal{S}(\hilbert_d)$ is a density matrix and $k$ is a real number. Thus, we have $k = \tr(\Delta)$. To extend the domain of $\mathcal{A}_{\mathcal{C}}^{\max}$ beyond density matrices and evaluate the assignment map on $\Delta$, it suffices to incorporate $k$ as the expectation value of the identity operator in~\eref{eq:PauliG}, introducing its own Lagrange multiplier, that is, $ k = \tr(\Id_{d} \Delta) = \tr(\Id_{d^n} \Delta_{\max}) = \tr(\Delta_{\max})$,
where $\Delta_{\max}$ is the maximum-entropy operator compatible with the normalization $k$ and with any other expectation value constrained by $\Delta$.

Since the identity operator commutes with all observables, it is easy to show that $\Delta_{\max} = k \varrho_{\max}$, where $\varrho_{\max}$ is the corresponding MaxEnt state compatible with $\Delta/k$. Therefore, the extension of $\mathcal{A}_{\mathcal{C}}^{\max}$ to Hermitian operators is homogeneous of degree 1, that is,
\[
\mathcal{A}_{\mathcal{C}}^{\max}(k\Delta) = k \mathcal{A}_{\mathcal{C}}^{\max}(\Delta)
\]
for any real scalar $k$. Consequently, $\mathcal{A}_{\mathcal{C}}^{\max}$ is trace preserving on this space. This also implies, trivially, that $\Gamma_t$ is trace preserving.

Provided that $\mathcal{A}_{\mathcal{C}}^{\max}$ and $\Gamma_t$ map density matrices to density matrices, the homogeneity of $\mathcal{A}_{\mathcal{C}}^{\max}$ implies the positivity of both maps. Indeed, for any $\Delta \geq 0$ we have
\[
\mathcal{A}_{\mathcal{C}}^{\max}(\Delta) = \left(\tr \Delta\right) \mathcal{A}_{\mathcal{C}}^{\max} \left( \frac{\Delta}{\tr \Delta} \right) \geq 0.
\]
By construction, $\mathcal{A}_{\mathcal{C}}^{\max}(\Delta/\tr \Delta)$ is a density matrix [see~\eref{eq:UnsimplifiedMaxEntState}]. Since both $\mathcal{C}$ and $\mathcal{V}_t$ are positive maps, it follows that $\Gamma_t$ is positive as well.
\subsubsection{Complete positivity of the assignment map} To prove that $\Gamma_t$ is completely positive, the crucial step is to show that $\mathcal{A}_{\mathcal{C}}^{\max}$ is completely positive. Due to the homogeneity of $\mathcal{A}_{\mathcal{C}}^{\max}$, it suffices to consider only density matrices in order to define its dilation. We denote this dilation by $\mathcal{A}^{\max}_{\mathcal{C} \otimes \text{id}_\text{E}}$ for any finite-dimensional ancillary system $\text{E}$. The dilation must satisfy the conditions
\begin{align}
\left(\mcC \otimes \text{id}_\text{E} \right) \circ  \mcA^\text{max}_{\mcC \otimes \text{id}_\text{E}} &= \text{id}_{\mcS(\hilbert_d \otimes \hilbert_\text{E})}, \nonumber \\
\mcA^\text{max}_\mcC(\rho) &=  \tr_\text{E} \mcA^\text{max}_{\mcC\otimes \text{id}_\text{E}}(\tilde\rho),
\label{eq:conditionsA_ext}
\end{align}
with $\rho = \tr_\text{E} \tilde \rho$. The first equation extends the consistency condition given in~\eref{eq:consistency}, while the second ensures that the effect of the dilation on the original system agrees with $\mathcal{A}^{\max}_\mathcal{C}(\rho)$.

To construct $\mathcal{A}^{\max}_{\mathcal{C} \otimes \text{id}_\text{E}}$, we apply the same procedure used to define $\mathcal{A}^{\max}_\mathcal{C}$, but now with $\mathcal{C} \otimes \text{id}_\text{E}$ as the coarse-graining map.

Consider a dilation of the microscopic description of the system with an
ancillary system described by the Hilbert space $\hilbert_\text{E}$; since the
extended coarse-graining map $\mcC\otimes \text{id}_\text{E}:\mcB(\hilbert_d^{\otimes n}\otimes \hilbert_\text{E}) \to \mcB(\hilbert_d\otimes \hilbert_\text{E})$ does not change
the dimension of the extension, we will use $\hilbert_\text{E}$ also for the
dilation of the coarse-grained description. Thus, the total state of the
extended microscopic
system is $\tilde \varrho\in \mcB(\hilbert_d^{\otimes n} \otimes \hilbert_\text{E})$. The next step is to extend the observables of the
coarse-grained description
$$\varsigma^\alpha\mapsto \varsigma^\alpha \otimes \Id_\text{E}$$
such that
$\langle \varsigma^\alpha \rangle
    =\tr{((\varsigma^\alpha \otimes \Id_\text{E}) \tilde \rho_\text{eff})}$
with $\tilde \rho_\text{eff}=\left(\mcC \otimes \text{id}_\text{E}
\right)(\tilde \varrho)$ the effective extended state.

Since the assignment map consists of preparing the maximum entropy state
compatible with the known expected values of $\varsigma_\alpha$, the following
expression holds:
$$\tr ((\varsigma^\alpha \otimes \Id_\text{E}) \tilde \rho_\text{eff})
          =\tr(\left(G^\alpha \otimes \Id_\text{E}\right) \tilde \varrho_{\max}). $$
The $G^\alpha$ is defined in \eref{eq:GOperators} and $\tilde \varrho_{\max}\in
\mcB(\hilbert_d \otimes \hilbert_\text{E})$ is the maximum entropy state
compatible with the mean values defined above. To construct explicitly
$\tilde \varrho_{\max}$ using the operators $G^\alpha \otimes \Id_\text{E}$, first
observe that $\exp\left(G^\alpha\otimes \Id_\text{E}\right)=
\exp\left(G^\alpha\right)\otimes \Id_\text{E}$; therefore, $\tilde \varrho_{\max}=\text{exp}\qty(\sum_{\alpha=1}^{d^{2}-1}\lambda_{\alpha}G^{\alpha}\otimes
\Id_\text{E})/\tilde Z=\exp\left(\sum_{\alpha=1}^{d^{2}-1}\lambda_{\alpha}G^{\alpha}\right)\otimes
\Id_\text{E}/\tilde Z$. Thus $\tilde Z=Z \times \dim\hilbert_\text{E}$ and the
$\lambda$s coincide with the ones in~\eref{eq:lambdas}. Therefore,
\begin{align}
\mcA^\text{max}_{\mcC \otimes \text{id}_\text{E}}(\tilde \rho_\text{eff})
  := \tilde\varrho_{\max}
     &= \varrho_{\max} \otimes \frac{\Id_\text{E}}{\dim \hilbert_\text{E}} \nonumber \\
     &= \mcA^\text{max}_{\mcC}(\rho_\text{eff})\otimes  \frac{\Id_\text{E}}{\dim \hilbert_\text{E}}
     \label{eq:extended_A}
\end{align}
holds, with $\rho_\text{eff}=\tr_\text{E}(\tilde\rho_\text{eff})$. This result is expected considering that the given expectation values $\langle \varsigma^\alpha \rangle$ do not give any information about the ancillary system. At this point we have proven that $\tilde\varrho_{\max}$ is a density matrix for any finite-dimensional ancillary system E. Moreover, the homogeneity of $\mcA^\text{max}_{\mcC \otimes \text{id}_\text{E}}$ on the space of Hermitian matrices implies complete positivity, i.e. $\mcA^\text{max}_{\mcC \otimes \text{id}_\text{E}}(\tilde \Delta) \geq 0$ for all $\tilde \Delta \geq 0$.

Observe that $\mcA^\text{max}_{\mcC \otimes \text{id}_\text{E}}$ is not written in the
usual fashion $\mcA^\text{max}_{\mcC} \otimes \text{id}_\text{E}$ as it is non linear,
but we were able to construct the extension anyway given that we can perform
the preparation of the maximum entropy state compatible with the local
observations, for any extension of the system. Moreover, notice that the formula in~\eref{eq:extended_A} trivially fulfills the conditions in~\eref{eq:conditionsA_ext}.

\subsubsection{Complete positivity of $\Gamma_t$} The composition of linear CPTP maps is trivially CPTP due to the semi group property of CPTP maps~\cite{Ziman2008}. However, in our case, a non linear map is involved. To prove that $\Gamma_t$ is CPTP, we compute its dilation based on the dilations of all constituent maps. The following scheme provides the details:
\[\begin{tikzcd}[arrows={<-|}]
    \tilde \rho_{\ef}(0)  & \left(\rho_{\ef}(t)\otimes \frac{\Id_\text{E}}{\dim \hilbert_\text{E}}\right) \arrow{l}{\tilde \Gamma_{t}} \arrow{d}{\mcC\otimes \text{id}_\text{E}}\\
   \left( \varrho_{\max}(0) \otimes \frac{\Id_\text{E}}{\dim \hilbert_\text{E}} \right) \arrow{u}{\mcA_{\mcC\otimes \text{id}_\text{E}}^{\max}} & \left(\varrho_{\max}(t) \otimes \frac{\Id_\text{E}}{\dim \hilbert_\text{E}}\right). \arrow{l}{\mcV_{t}\otimes \text{id}_\text{E}}
\end{tikzcd}
\]
Thus, the dilation of $\Gamma_t$ results in
\begin{equation}
    \begin{gathered}
        \tilde \Gamma_{t}: \mcS( \hilbert_{d}\otimes \hilbert_\text{E})\rightarrow\mcS( \hilbert_{d} \otimes \hilbert_\text{E}),\\
       \tilde \rho_{\ef}(0) \mapsto \Gamma_t(\rho_\text{eff}(0))\otimes \frac{\Id_\text{E}}{\dim \hilbert_\text{E}},
    \end{gathered}
\end{equation}
where $\rho_\ef(0) = \tr_\text{E} \tilde{\rho}_\ef(0)$. Therefore, $\tilde{\Gamma}_{t}(\tilde{\rho})$ is a density matrix for all $\tilde{\rho} \in \mathcal{B}(\hilbert_d \otimes \hilbert_\text{E})$. This, combined with the homogeneity of the extension of $\tilde{\Gamma}_{t}$ to the space of Hermitian matrices (inherited from $\mathcal{A}^\text{max}_{\mathcal{C} \otimes \text{id}_\text{E}}$), implies
\begin{equation}
\tilde{\Gamma}_{t}(\tilde{\Delta}) \geq 0 \, \forall \tilde{\Delta} \in \mathcal{B}(\hilbert_d \otimes \hilbert_\text{E}), \quad \tilde{\Delta} \geq 0.
\end{equation}
In conclusion, $\Gamma_t$ is a non linear CPTP map.

\section{Effective non-linear quantum gates} \label{sec:EffectiveGates} In light of the framework we have developed, and before examining
the effective dynamics of $n$-qubit systems, we will focus our
attention on simpler two-qubit systems, where the effect of the coarse-graining
map on a microscopic state $\varrho$ will be
\begin{equation}\label{eq:CGTwoQubit}
  \mcC(\varrho)=p_{1}\tr_{2}(\varrho)+p_{2}\tr_{1}(\varrho),
\end{equation}
which is a convex combination of the reduced states of the two qubits and is equivalent
to \eref{eq:CG} for $n=2$.
In particular, we will explore 
the effective dynamics generated by two well known quantum gates: the 
\textsc{swap} gate and the controlled-\textsc{not} (\textsc{cnot}) gate. We will see that the 
resulting effective dynamics is far from uninteresting. Indeed, the effective
dynamics induced by these gates exhibit non linearity and non-Markovian
behavior \cite{NonMarkovian}. 
\subsection{Effective SWAP gate} 
\begin{figure}[ht!]
  \hspace{-8mm}
  {\includegraphics{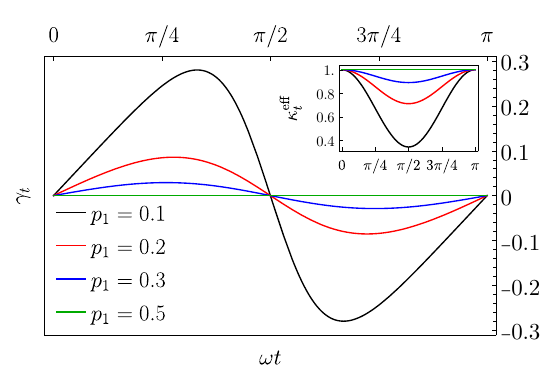}}
  \caption{\label{fig:K(t)}
    Plot of the decay rate $\gamma_t$ and the polarization coefficient $\kappa_t^{\ef}$ (inset) for different values of $p_1$, using an initial effective state with $r_{\ef} = 0.9$. The decay rate exhibits both positive and negative values, indicating non-Markovian behavior in the dynamics.}
\end{figure}

The Hamiltonian $H_\swap$ given by
\begin{equation}\label{eq:SWAPHamiltonian}
  H_{\swap}=\frac{\omega}{2}\sum_{\alpha\in\{x,y,z\}}\sigma^{\alpha}\otimes\sigma^{\alpha}
\end{equation}
has the property of generating the $\swap$ gate for
$t={\pi}/{2\omega}$. 
Following the definition of $\Gamma_{t}$ [\eref{eq:GammaMap}], we find that the effective state of the evolved 
system is given by
\begin{equation}
  \begin{aligned}
    \mcC(U_{\swap}(t)\varrho_{\max}U_{\swap}^{\dagger}(t))=&\qty(p_{1}\sin^{2}\omega t+p_{2}\cos^{2}\omega t)\varrho_{1}\\
    +&\qty(p_{1}\cos^{2}\omega t+p_{2}\sin^{2}\omega t)\varrho_{2},
  \end{aligned}\nonumber
\end{equation}
where $U_{\swap}(t)$ is the unitary operator generated by the Hamiltonian \eqref{eq:SWAPHamiltonian}. 
Since the effective initial state is given by~\eref{eq:CGTwoQubit}, it is
possible to see that the effective state's Bloch vector has a constant direction; only  
its length changes with time. This means that the effective dynamics
corresponds to a depolarization channel where the depolarization coefficient depends on 
the initial state. To facilitate our analysis, let us define the Bloch vector's norm 
for the $k$th particle as $r_{k}=\tanh{(p_{k}\lambda)}$ and similarly denote the
norm of the Bloch vector for the effective state by $r_{\ef}$. With this, the quotient 
$\kappa_t^{\ef}$ between the norm of the Bloch vector of the initial effective state 
and the norm of the Bloch vector of the effective state at time $t$ is given by
\begin{equation}
  \begin{aligned}\label{eq:KappaSWAP}
  \kappa_{t}^{\ef}=\frac{1}{r_{\ef}(0)}\big[&p_{1}(r_{1}\cos^{2}{\omega t}
  +r_{2}\sin^{2}{\omega t})\\
  &+p_{2}(r_{2}\cos^{2}{\omega t}+r_{1}\sin^{2}{\omega t})\big].
\end{aligned}
\end{equation}
This allows us to write the effective dynamics of the system as
\begin{equation}\label{eq:EffectiveSWAPt}
  \Gamma_{t}(\rho_{\ef})=\kappa_{t}^{\ef}\rho_{\ef}+(1-\kappa_{t}^{\ef})\frac{1}{2}\Id.
\end{equation}
Furthermore, since the effective dynamics is a depolarization channel,
the evolution of the effective state is described by the Lindblad-like
differential equation
\begin{equation}\label{eq:SWAPDiffEq}
  \dot{\rho}_{\ef} = \gamma_t
  \sum_{\alpha\in\{x,y,z\}}\qty(\sigma^{\alpha} \rho_{\ef} \sigma^{\alpha} - \rho_{\ef}),
\end{equation}
with $\gamma_t=-\dot{\kappa}_{t}^{\ef}/4\kappa_{t}^{\ef}$.
A plot of the decay rate and the depolarization coefficient is shown in 
\fref{fig:K(t)}. This reveals that there are some scenarios where the effective
state remains unchanged over time. For example, in the non preferential case where
$p_{1}=p_{2}$, the microscopic state is invariant under the \textsc{swap} gate 
[see~\eref{eq:NonPreferentialAss}], and therefore the effective state remains invariant
under the effective dynamics. In fact, it is easy to see that the depolarization 
coefficient \eqref{eq:KappaSWAP} is equal to one when $p_{1}=p_{2}$ 
(which is also shown in~\fref{fig:K(t)}).
Additionally, all pure states $\ketbra{\phi}$ are also invariant under 
the effective dynamics, as long as no $p_{k}=0$, since the only 
microscopic states compatible with such a macroscopic state is $\ketbra{\phi}^{\otimes n}$,
a symmetric state.
Most interesting, is that the decay rate $\gamma_t$ is able to
take negative values, indicating that, in addition to being non linear,
the effective dynamics is non-Markovian~\cite{Breuer,davalosprl,montesgorin,davalosdiv}. Quantitatively, using the Breuer-Laine-Piilo 
measure in the interval shown in~\fref{fig:K(t)}, we have~\cite{blpmeasure}
\begin{align}
\mcM_\text{BLP} [\Gamma_t] &= \int_{\dot k^{\ef}_t>0}\dot k^{\ef}_t dt \nonumber\\
&= k^{\ef}_{\pi/\omega} - k^{\ef}_{\pi/2\omega}  \nonumber\\
&= \frac{(p_1-p_2)(r_1-r_2)}{r_\text{eff}(0)}\geq0.
\end{align}
The expression in the last line is always non-negative because \( p_2 > p_1 \) implies \( r_2 > r_1 \), and vice versa, since \( \tanh \) is strictly increasing over the reals 
[see~\eref{eq:LambdaNorm}]. 

To perform the optimization required to compute \( \mathcal{M}_\text{BLP} \), we used the states \( \dyad{0} \) and \( \dyad{1} \); however, any pair of antipodal states on the Bloch sphere yields maximal distinguishability. This is because \( \Gamma_t \) is a non linear depolarizing channel that compresses or inflates the Bloch sphere in an isotropic manner.

\subsection{Effective CNOT gate} Similarly to \eqref{eq:SWAPHamiltonian}, the Hamiltonian
\begin{equation}
  H_{\cnot}=-\frac{\omega}{2}(\sigma^{z}\otimes\Id+\Id\otimes\sigma^{x}-\sigma^{z}\otimes\sigma^{x}),
\end{equation}
generates the \textsc{cnot} gate at time $t=\frac{\pi}{2\omega}$. The effective
dynamics at an arbitrary time $t$ might be found by following the same 
procedure as in the previous case. However, the resulting expression is
cumbersome and not very illuminating. Instead, we focus on the
effective dynamics of the system at time $t=\frac{\pi}{2\omega}$,
which is the time at which the \textsc{cnot} gate is generated. The effective
dynamics at this time is given by
\begin{equation}\label{eq:EffectiveCNOT}
  \begin{aligned}
  \Gamma_{\cnot}(\rho_{\ef})=
  \frac{1}{2}\big[\rho_{\ef}&+p_{1}\mcD_{\expval{\sigma_{2}^{x} }}^{z}(\rho_{1})\\
  &+p_{2}\mcD_{\expval{\sigma_{1}^{z} }}^{x}(\rho_{2})\big],
\end{aligned}
\end{equation}
where $\mcD_{q}^{\beta}$ is a dephasing channel along 
the $\beta$ direction with dephasing coefficient $1-q$. For example, if $\beta=z$, then
\begin{equation}\label{eq:dephasing_channel}
  \mcD_{q}^{z}(\rho) = q\rho+(1-q)\sigma^{z}\rho\sigma^{z};
\end{equation}
that is, in \eqref{eq:EffectiveCNOT} we encounter a convex
combination of the effective state and two non linear state-dependent
dephasing channels. Note that the usual interpretation of the \textsc{cnot} gate
is recovered, as we see that a bit flip (phase flip) is applied to the
second (first) particle depending on the state of the first (second)
particle. At an arbitrary time $t$, the effective dynamics is given by
\begin{equation}
  \Gamma_{t}(\rho_{\ef})=
  \frac{1}{2}\rho_{\ef}+\frac{p_{1}}{2}\mcE^{x}(\rho_{\ef})+\frac{p_{2}}{2}\mcE^{z}(\rho_{\ef}),
\end{equation}
where
\begin{align}
  \mcE^{x}(\rho_{\ef})= & \rho_{1}\cos^{2}(\omega t)\nonumber \\
  & +\qty[\expval{\sigma^{x}_{2}}\rho_{1}+
  \qty(1-\expval{\sigma^{x}_{2}})\sigma^{z}\rho_{1}\sigma^{z}]\sin^{2}(\omega t)\nonumber \\
  & -i\qty(1-\expval{\sigma^{x}_{2}})\cos(\omega t)\sin(\omega t)[\rho_{1},\sigma^{z}]\nonumber,
\end{align}
and
\begin{align}
  \mcE^{z}(\rho_{\ef})= & \rho_{2}\cos^{2}(\omega t)\nonumber\\
  & +\qty[\expval{\sigma^{z}_{1}}\rho_{2}+
  \qty(1-\expval{\sigma^{z}_{1}})\sigma^{x}\rho_{2}\sigma^{x}]\sin^{2}(\omega t)\nonumber \\
  & -i\qty(1-\expval{\sigma^{z}_{1}})\cos(\omega t)\sin(\omega t)[\rho_{2},\sigma^{x}]\nonumber.
\end{align}
In Appendix \ref{ap:EllipticCNOT} we show that $\mcE^{z}$ and $\mcE^{x}$
are channels that describe elliptical trajectories on the Bloch sphere. 
This is because they correspond to the reduced dynamics of two 
two-level systems that evolve according to a non local Hamiltonian \cite{EllipticalOrbits}.

\section{Effective spin chain dynamics} \label{sec:SpinChains} Now that we have seen that non linear, non-Markovian results arise
from the coarse-grained dynamics of a two-qubit system, we study
larger many-body systems, such as the spin chains. 
\subsection{Non uniform external magnetic field with all-to-all interactions} First, we consider spin-$1/2$ chains evolving in a non uniform external
magnetic field in the $z$ direction with an all-to-all Ising interaction
parallel to the field.  The Hamiltonian considered is thus
\begin{equation}\label{eq:NonUniformField}
    H = H_{\text{field}}+H_{\text{int}},
\end{equation}
where
\begin{equation}
    H_{\text{field}}=\sum_{k}\omega_{k}\pauli{k}^{z},\quad
H_{\text{int}}=(\sigma^{z})^{\otimes n}. 
\end{equation}
Since the field and interaction terms commute, it is possible to solve
them separately. \par

Let us first consider the magnetic-field part of the Hamiltonian,
which leads to an
effective evolution that is a combination of all the local evolutions.
Indeed, because all terms in $H_{\text{field}}$
commute, the corresponding unitary evolution is
\begin{align}\label{eq:Ufield}
    U_{\text{field}}(t)=\prod_{k}U_{k}(t),
\end{align}
where $U_{k}(t)=e^{-i\omega_{k}t\pauli{k}^{z}}$. 
Applying the coarse-graining map to the maximum entropy stated evolved through
\eqref{eq:Ufield} yields
\begin{equation}\label{eq:NonUniformFieldEffectiveEvolution}
    \Gamma_{t}^{\text{field}}(\rho_{\ef})=\sum_{k}p_{k}U_{k}(t)\rho_{k}U^{\dagger}_{k}(t),
\end{equation}
where $\rho_{k}$ are the subsystems
defined by the maximum entropy assignment map \eqref{eq:MaxEntAss}. \par

\begin{figure} \includegraphics[width=0.48\columnwidth]{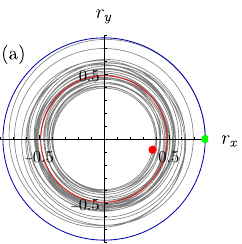}
\includegraphics[width=0.48\columnwidth]{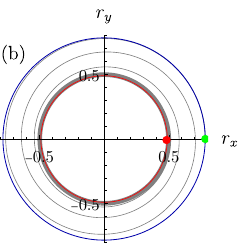}
\quad
\caption{\label{fig:EffLocalNoPref} Effective evolution of the
macroscopic state's Bloch vector for (a) $n=10$ and (b) $n=500$ particles
under the Hamiltonian $H_\text{field}$ with frequencies $\omega_{k}$ 
normally distributed with mean $\mu = 1.5$ and standard deviation $\varsigma = 0.2$ (manifested in the variation of the gray curve around the limit cycle). The green point represents the state at $t=0$ and the red
point denotes the state at the final time $t=4t_c$. The effective state oscillates
around a mean evolution (red) of radius $p_{1}r_1$, with $p_1=0.5$.
Comparing the two plots, its possible to appreciate the decrease of the
fluctuations, which is proportional to $1/\sqrt{n}$.}
\end{figure}

We now show that in scenarios where a dominant probability prevails, the
state evolution described by Eq.~\eqref{eq:NonUniformFieldEffectiveEvolution}
tends to spiral down towards a
unitary evolution into a more mixed state. Specifically, each term in the
summation of Eq.~\eqref{eq:NonUniformFieldEffectiveEvolution} circles the
$z$ axis. However, given random frequencies, these movements are incoherent.
This lack of coherence manifests completely after a specific time, defined as
$t_c = 2 \pi/ \varsigma$, where $\varsigma$ represents the standard deviation
of the frequencies $\omega_k$. At this juncture, under appropriate conditions,
the phases $\exp(i \omega_k t)$ distribute uniformly around the whole unit
circle. In scenarios where one probability, such as $p_1$, significantly
surpasses others and there is a large number of particles, the dominant effect
is a singular weakened contribution
\begin{align}\label{eq:FieldConvergentEvolution}
\lim_{\substack{n \to \infty \\ t > t_c}}
\Gamma^{\text{field} }_t (\rho_{\ef}) \to
U_{1}(t)\mcP_{p_1r_1}(\rho_{\ef})U_{1}^{\dagger}(t),
\end{align}
where $\mcP_{q}(\rho)=q\rho+(1-q)\Id/d$ is a depolarizing channel. 
Thus, after $t_c$, the dynamics converges to a limit cycle in the Bloch 
sphere representation [see \fref{fig:EffLocalNoPref}(a)]. 
In fact, the fluctuations around this evolution will decrease as 
approximately $1/\sqrt{n}$, since we would
be effectively adding $n-1$ random numbers on top of the preferred evolution
[see~\fref{fig:EffLocalNoPref}(b)]. 
This convergence reflects how dominant probabilities influence the system's
evolution, simplifying the overall dynamics to primarily one major
contribution.
The contraction of the resulting state with respect to the initial one is a
consequence of the loss of information about the non preferential particles.
This loss of information arises due to the averaging process of the
preferential measurement, which effectively discards information about the
other particles and makes the effective state more mixed.
It is worth adding that if there is no preferential particle, we will observe 
the qubit simply spiraling towards the origin in the $x$-$y$ plane while keeping the $z$ component constant. \par

 Having solved the free part of the Hamiltonian, we now focus on the
interaction term.  It can be shown, either by applying the coarse-graining map
to the evolved microscopic state or by applying it to the Liouville--von Neumann
equation, that the interaction part of the Hamiltonian leads approximately to a
non unitary but linear evolution. Consider the Liouville--von Neumann equation
\begin{equation}
    \dot{\varrho}_{\max}=-i[(\sigma^{z})^{\otimes n},\varrho_{\max}]
\end{equation}
and solve it by iteratively integrating and substituting the implicit solution in the commutator,
from which we obtain the power series in $H_{\text{int}}$. Applying the coarse-graining map and
taking into account its linearity, we arrive at
\begin{equation}\label{eq:Dyson-like}
    \begin{aligned}
    \Gamma^{\text{int}}_{t}(\rho_{\eff})= & \rho_{\eff}(0)+(-it)\mcC\left([H_{\text{int}},\varrho_{\max}(0)]\right)                                      \\
                                          & +\frac{(-it)^2}{2!}\mcC\left([H_{\text{int}},[H_{\text{int}},\varrho_{\max}(0)]]\right)                      \\
                                          & +\cdots .
    \end{aligned}
\end{equation}
To work with~\eref{eq:Dyson-like}, we notice that the
nested commutators can be expressed as
\begin{widetext}
    \begin{equation}\label{eq:CommutatorCases}
        [\smash[b]{H_{\text{int}}\kern-2pt
        \underbracket[0.5pt][1pt]{, \ldots,[H_{\text{int}},}_\text{$n$ times}}
        \varrho_{\max}(0)]] 
        = \begin{cases}
            2^{n-1}[H_{\text{int}},\varrho_{\max}(0)], & \text{odd $n$} \\
            2^{n-1}[\varrho_{\max}(0)-H_{\text{int}}\varrho_{\max}(0) H_{\text{int}}], & \text{even $n$}.
        \end{cases}
    \end{equation}
\end{widetext}
Using this, we can integrate the von Neumann equation as usual to obtain two
different power series, corresponding to even and odd powers of $t$.
Then we can neglect the terms with odd powers of $t$ in the limit
of large $N$, because $\mcC\{[H_\text{int},\varrho_{\max}(0)]\}$
decreases exponentially with  the number of particles.
This will lead us to find that the effective evolution for big $N$ is
\begin{equation}\label{eq:EffectiveDephasing}
    \Gamma_{t}^{\text{int}}(\rho_{\ef})=\mcD_{\cos^{2}\qty(t)}^{z}(\rho_{\ef}),
\end{equation}
i.e., a dephasing channel in the $z$ direction with oscillating intensity. For
details regarding the calculation stated in this paragraph, see 
Appendix~\ref{ap:small_commutator}. \par

\begin{figure} \includegraphics[width=0.48\columnwidth]{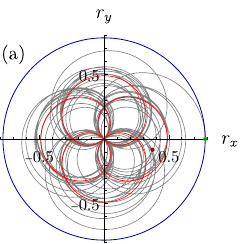}
    \includegraphics[width=0.48\columnwidth]{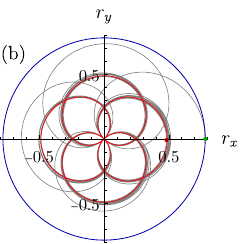}
    \quad
    \caption{\label{fig:EffLocalIIntNoPref} 
    Plot of the effective evolution of a set of particles, under the Hamiltonian
    \eref{eq:NonUniformField} for  (a) $n=10$ and (b) $n=500$ particles. 
    See the caption of \fref{fig:EffLocalNoPref} for the color coding. The frequencies
    are chosen in an identical manner, as well as the probability distribution and the evolution time. 
    In this case, the asymptotic evolution, in red, corresponds to \eref{eq:EffectivePolar}.
}
\end{figure} 

Now that we have found the solutions to both the local and interaction
parts of the Hamiltonian to be given by \eqref{eq:NonUniformFieldEffectiveEvolution}
and \eqref{eq:EffectiveDephasing}, respectively, and given that both dynamics commute,
we can say that the effective
evolution will be the composition of these two, that is,
\begin{align}\label{eq:EffectivePolar}
    \lim_{\substack{n \to \infty \\ t > t_c}}
    \Gamma_t (\rho_{\ef}) \to
    \mcD_{\cos^{2}\qty(t)}^{z}\left(U_{1}(t)\mcP_{p_1r_1}(\rho_{\ef})U_{1}^{\dagger}(t)\right).
\end{align}
For times much larger than $t_{c}$, 
the effective state's Bloch vector will oscillate around
the composition of the evolution of the preferred subsystem and the dephasing
channel, i.e., a polar rose,  as shown in \fref{fig:EffLocalIIntNoPref}.
In cylindrical coordinates, we can write 
\begin{align}
    r(t)&=r(0)\cos[\theta(t)], \\
    \theta(t)&=\frac{t}{\omega_{1}} +\theta(0),\\
    z(t)&=z(0).
\end{align}

It is important to underline that such dynamics is obtained assuming a large
number of particles. This allows us to write the interaction part of the
evolution as a dephasing channel, i.e., as a straight line. For finite $n$, a
squeezed elliptical trajectory is found. Moreover, the actual trajectory will
oscillate around \eref{eq:EffectivePolar} with an amplitude proportional to
$1/\sqrt{n}$.

So far, our discussion of the dynamics induced by Hamiltonian \eqref{eq:NonUniformField}
has centered around the use of the preferential distribution \eqref{eq:ProbDistributions}.
For the non preferential case, the situation is analogous. Indeed, without interaction, the
dynamics will once again follow \eqref{eq:FieldConvergentEvolution}. However, since the
depolarization of the state will have a coefficient $r_\text{eff}/n$, it will effectively converge to
a point on the $z$ axis. On the other hand, for the interacting case, the state's trajectory 
will follow the shape of the appropriate rose, but with an increasingly small amplitude, until
it also collapses to the $z$ axis.
\subsection{Ising chain} We now consider a homogeneous and closed Ising spin-$1/2$ chain coupled to a
transversal magnetic field. The governing Hamiltonian is
\begin{equation}\label{eq:IsingHamiltonian}
    H=-J\sum_{j=1}^N \pauli{j}^{z}\pauli{j+1}^{z}-g\sum_{j=1}^N\pauli{j}^{x}, \quad
\sigma^z_{N+1} \equiv \sigma^z_1.
\end{equation}
This Hamiltonian is invariant under spin translations, that is, changing
indices $j\to j+1$.
We study the case when the effective initial state is pure,
$$\ket{\psi}=\cos\left( \frac{\theta}{2} \right)\ket{0}+ie^{\i \phi}\sin\left( \frac{\theta}{2} \right)\ket{1}.$$
When all particles participate in the coarse graining, i.e., they have non zero
probabilities in \eref{eq:CG}, the only compatible microscopic state is
$\ket{\psi}^{\otimes N}$, given that pure states are extremal. 

Moreover, observe that the reduced dynamics of one spin is the same
for every spin due to the translation symmetry of the Hamiltonian and
the permutation symmetry of the initial state. If we denote
the reduced density matrix of the
$k$th particle at a time $t$ by $\rho_k(t)$,  
this implies that $\rho_k(t)=\rho_l(t)$ for every
$t$, $k$, and $l$. Consequently,
the effective state at time $t$ is simply
$\rho_\eff(t)=\rho_k(t)$ for any $k$. Therefore, 
the effective dynamics and
the microscopic state are independent of the exact values of
$p_k$, provided all of them are
non zero ($p_k>0 \, \forall \, k$).

\begin{figure}
\centering
\includegraphics[width=\columnwidth]{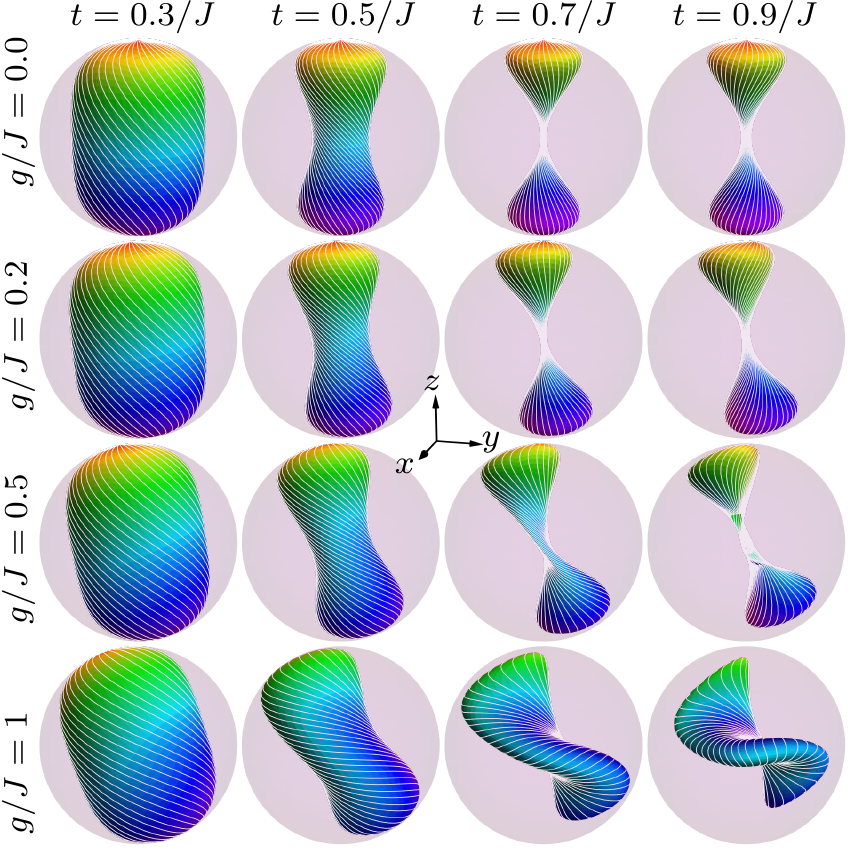}
\caption{Evolution of the Bloch sphere containing all initial pure states for several values of field coupling and time, with eight qubits in the fine transverse-field Ising dynamics. The color code is as follows: A gradient from north (red) to south (blue) represents different polar angles, while white wires indicate families of states with a fixed azimuthal angle.  The effective dynamics is nonlinear in all cases; in particular, for $g/J=0$, it corresponds to nonlinear dephasing with differential rotation around the $z$ axis (illustrated by the twisted white wires). See main text for further details.
\label{fig:ising_1}}
\end{figure}

Let us now discuss the results. For $g/J=0$ the system is tractable
analytically (see Appendix~\ref{app:ising} for details). For this case,
the reduced dynamics is a non linear dephasing with rotation around the $z$
axis,
\begin{equation}
\begin{aligned}
&\rho_\eff(t)=\\
&\left(
\begin{array}{cc}
 \cos ^2\left(\frac{\theta }{2}\right) & \gamma(\theta,t)\frac{1}{2} e^{-i \phi } \sin (\theta ) \\
 \gamma^{*}(\theta,t)\frac{1}{2} e^{i \phi } \sin (\theta ) & \sin ^2\left(\frac{\theta }{2}\right) \\
\end{array}
\right),
\end{aligned}
\end{equation}
with
\begin{equation}
\gamma(\theta, t)=\left[\cos\left(2J t\right)+i \cos(\theta)\sin\left(2J t\right)\right]^2.
\label{eq:gamma_ising}
\end{equation}
The dynamics is non linear and does not depend on the total number of spins in
the chain. Moreover, it depends only on $\theta$ (see~\fref{fig:ising_1}); this
is expected due to the azimuthal symmetry of the Hamiltonian. Furthermore, both
dephasing and rotation depend on $\theta$; thereforem, there is a differential
rotation apart from the non linear dephasing, depicted by the twisted white 
wires in the figures.

For $g/J > 0$, we performed numerical
calculations without approximations for $2252$ initial pure states, uniformly
distributed to resemble the polygon mesh representation of the Bloch sphere.
In~\fref{fig:ising_1} we present the results for $N = 8$, where it can be
observed that the dynamics becomes increasingly intricate as $g/J$ increases.

In our numerical experiments, we observed that the results are visually
indistinguishable for $N \geq 4$. Since increasing the system's size is
computationally expensive, we analyzed how uniformly random effective 
initial states vary with respect to the finer system's size. To do this, 
we evolved each random state over multiple time points, particle numbers 
at the finer level, and different several values of $g/J$. 

\begin{figure}
\centering
\includegraphics[width=\columnwidth]{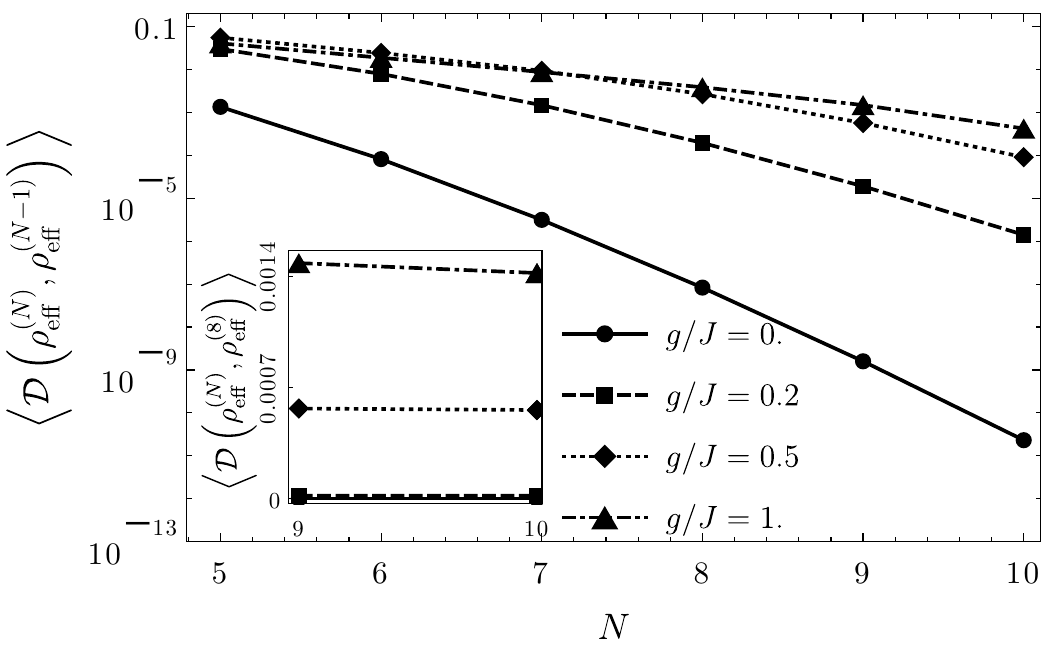}
\caption{Average trace distance of effective states evolved with different numbers of particles at a finer level. The main plot shows the average trace distance between consecutive particle numbers. The inset displays the average trace distance for nine and ten particles with respect to the case of eight particles. The average is taken over multiple time points and 100 quantum states. \label{fig:ising_error}}
\end{figure}

For comparison, in~\fref{fig:ising_error} we plot the trace distance 
between states evolved with consecutive particle numbers, while the 
inset shows the trace distance with respect to the case of eight particles 
(recalling that~\fref{fig:ising_1} was produced with eight particles). 
As $g/J$ increases, the error grows slightly but still decays 
rapidly. With respect to the system with eight particles 
(inset in~\fref{fig:ising_error}), the error becomes 
negligible, indicating that the results shown in~\fref{fig:ising_1} 
are close to those of the 
thermodynamic limit ($N \to \infty$).

\section{Linear effective dynamics}\label{sec:LinearDynamics} In~\sref{sec:EffectiveGates} we presented two examples
of microscopic dynamics that give rise to a non linear, non-Markovian
evolution of the effective state. Nevertheless, not all microscopic
dynamics will produce such behavior of the effective dynamics. In this
section we present some examples of evolution for which the linearity
is preserved under the coarse-graining map. 
\subsection{Channels that act equally on all subsystems} 

First, let $\mcE$ be a quantum channel acting on a two-level system
\begin{equation}\label{eq:channel}
    \mcE:\mcB(\hilbert_{2}) \rightarrow \mcB(\hilbert_{2})
\end{equation}
and let $\varrho$ be the density operator that describes a
collection of $n$ such units. We define $\mcE^{\otimes n}$ as 
$\mcE$ applied locally to each of the subsystems of the 
$n$-partite system,
\begin{equation}\label{eq:channelN}
    \mcE^{\otimes n}:\mcB(\hilbert_{2}^{\otimes n}) \rightarrow \mcB(\hilbert_{2}^{\otimes n})\rlap{,}
\end{equation}
according to
\begin{equation}
    (\mcE\otimes\mcE)(A \otimes B)=\mcE(A)\otimes\mcE(B)\rlap{.}
\end{equation}
for any $A,B\in\mcB(\hilbert_{2})$. By expanding 
$\varrho$ in the basis of tensor products of the
Pauli matrices it is a matter of algebra to show
that
\begin{equation}
    \qty(\mcC\circ\mcE^{\otimes n}) (\varrho) =\qty(\mcE\circ\mcC)(\varrho)
\end{equation}
for any $\varrho\in\mcS \qty(\hilbert_{2}^{\otimes n})$. 
This means that all channels that act locally and
in the same way over all subsystems conserve their 
linearity under the coarse-graining map. Furthermore,
non factorizable channels that act equally upon all
reduced matrices also conserve their linearity.

As a first example, let us consider dephasing channel. 
An $n$-qubit total dephasing channel in the $z$ direction is
\begin{equation}
    \mcD^{z}_{1/2^{n}}(\varrho)=\frac{1}{2^{n}}\sum_{\vec{\alpha}}(\sigma^{\vec{\alpha}})\varrho (\sigma^{\vec{\alpha}})^{\dag},
\end{equation}
where $\sigma^{\vec{\alpha}}=\sigma^{\alpha_{1}}\otimes\cdots\otimes\sigma^{\alpha_{n}}$ with $\alpha_{j}\in\{0,z\}$. It is not difficult to see that
\begin{equation}
    \mcC\qty(\mcD^{z}_{1/2^{n}}(\varrho))=\mcD^{z}_{1/2}\qty(\mcC\qty(\varrho)),
\end{equation}
that is, the total dephasing channel on the microscopic system
translates under the coarse-graining map as a total dephasing channel
over the effective system.
The total dephasing channel is an example of a factorizable Pauli component erasing map \cite{PCE}.
A non factorizable Pauli component erasing map that is also a quantum
channel is 
\begin{equation}
    \begin{gathered}
        \mcE:\mcB(\hilbert_{2}^2) \rightarrow \mcB(\hilbert_{2}^2),\\
        \varrho=\frac{1}{4}\sum_{\vec{\alpha}}\gamma_{\vec{\alpha}}\sigma^{\vec{\alpha}}\rightarrow\frac{1}{4}\sum_{\vec{\alpha}}\gamma_{\vec{\alpha}}\tau_{\vec{\alpha}}\sigma^{\vec{\alpha}},
    \end{gathered}
\end{equation}
with
\begin{equation}
    \begin{pmatrix}
        \tau_{0,0} & \tau_{0,x} & \tau_{0,y} & \tau_{0,z} \\
        \tau_{x,0} & \tau_{x,x} & \tau_{x,y} & \tau_{x,z} \\
        \tau_{y,0} & \tau_{y,x} & \tau_{y,y} & \tau_{y,z} \\
        \tau_{z,0} & \tau_{z,x} & \tau_{z,y} & \tau_{z,z}
    \end{pmatrix}
    =
    \begin{pmatrix}
        1 & 0 & 1 & 0 \\
        0 & 1 & 0 & 1 \\
        1 & 0 & 1 & 0 \\
        0 & 1 & 0 & 1
    \end{pmatrix}.
\end{equation}
Note that this $\mcE$ acts equally on both reduced
density matrices. In fact, it acts as a dephasing 
channel in the $y$ direction upon both subsystems,
and so
\begin{equation}
    \mcC\qty(\mcE(\varrho))=\mcD^{y}_{1/2}\qty(\mcC\qty(\varrho)).
\end{equation}
Although this is an example for two two-level
systems, all $n$-qubit quantum channels that satisfy
\begin{equation}\label{eq:condition}
    \tr_{\overline{j}}(\mcE(\varrho))=\tr_{\overline{k}}(\mcE(\varrho))\, \forall \, k,j\in\{1,\ldots,n\}
\end{equation}
will give rise to linear effective dynamics.

\subsection{A linear non-markovian evolution} \label{subsec:nonmarko}
Another instance of a microscopic evolution that preserves its
linearity under the coarse-graining map is the given by the
Hamiltonian
\begin{equation}\label{eq:linear_non_markovian}
  H = \frac{\omega}{2}(\Id\otimes\sigma^z).
\end{equation}
Clearly, this is not a mapping that abides by the  condition stated 
by~\eref{eq:condition}. However, as we will see, the effective
dynamics remain linear with a notable twist: They are non-Markovian.
Indeed, using the non preferential distribution 
described~\eref{eq:ProbDistributions}, the microscopic evolution 
\eqref{eq:linear_non_markovian} induces the macroscopic
dynamics given by
\begin{equation}\label{eq: eff dynamics for first local case of 2 particles}
  \Gamma_t(\rho_\ef) = \frac{1}{2}\qty(\rho_\ef + e^{-i (\omega t /2)\sigma^z}\rho_\ef e^{i (\omega t /2)\sigma^z}),
\end{equation}
and its effect on the evolution of the effective Bloch vector is
described by the parametric equations
\begin{align}
  r^x_{\eff}(t) & = \frac{1}{2}\qty[r^x_{\eff} \cos(\omega t) - r^y_{\eff}\sin(\omega t) + r^x_{\eff}] \\
  r^y_{\eff}(t) & = \frac{1}{2}\qty[r^y_{\eff}\cos(\omega t) + r^x_{\eff}\sin(\omega t) + r^y_{\eff}]  \\
  r^z_{\eff}(t) & = r^z_{\eff},
\end{align}
which represent the parametric equations of a circle in $\mathbb{R}^3$ with
center at $(r^x_{\eff}/2, r^y_{\eff}/2, r^z_{\eff})$, parallel to the $x$-$y$ plane, and
with radius
$\frac{1}{2}\sqrt{(r^x_{\eff})^2 + (r^y_{\eff})^2}$

The differential equation governing the effective dynamics is
\begin{equation}\label{eq: diff equation for first local case of 2 particles}
  \dv{\rho}{t} = -i\qty[\frac{\omega}{4}\sigma^z, \rho] + \frac{\omega}{2}\tan(\frac{\omega t}{2})(\sigma^z \rho \sigma^z - \rho).
\end{equation}
From this differential equation, we can identify the effective Hamiltonian as
\begin{equation}
  H_\ef = \frac{\omega}{4}\sigma^z,
\end{equation}
where the frequency appearing in the effective Hamiltonian is reduced by half
compared to the frequency of the microscopic Hamiltonian.  It is worth
noting that~\eref{eq: diff equation for first local case of 2
particles} exhibits singularities at points where the tangent function
diverges.

Since the term $\tan(\frac{\omega t}{2})$ can be positive or negative, the
dynamics is generally non-Markovian. Furthermore, it is straightforward to
verify that the corresponding dynamics in \eqref{eq: eff dynamics for first
local case of 2 particles} do not satisfy the semigroup property: 
\begin{equation}
  \Gamma_{t+s}(\rho) \neq \Gamma_t(\Gamma_s(\rho)).
\end{equation}
Thus, coarse-grained quantum dynamics adds to the wide variety of physical systems 
that exhibit non-Markovianity~\cite{LI20181}.

\section{Discussion and conclusions \label{sec:conclusions}} 

We introduced a pragmatic notion of
effective dynamics for coarse-grained descriptions of conventional
many-body quantum systems. To achieve this, we used the maximum
entropy principle to deal with the fact that the effective description emerges
from a coarse-graining map that irreversibly destroys part of the microscopic
information. This method is quite appealing given that no
additional information or assumptions are introduced to construct the microscopic
state of the system. We assumed that the latter undergoes unitary
evolution and thus the system is closed. Remarkably, the emergent dynamics is
generally non linear and depends on both the initial effective state and the
probability distribution that defines the coarse-graining map. 

Additionally, we have proven that such dynamics defines non linear completely
positive and trace preserving-maps. We believe that this is a beneficial trait
given that complete positivity implies positivity, and classical stochastic
maps are positive~\cite{Breuer}. Moreover, in the absence of quantum
correlations, complete positivity is reduced to positivity~\cite{holevo}.
This ensures consistency if we consider this framework to investigate the
quantum-to-classical transition.

We studied several systems to test the framework, ranging from quantum gates to
spin systems. They show explicitly the dependence on the distribution of the
coarse-graining map and on the initial effective states. However, in the case
of the Ising spin chain, its symmetries conceal the effective dynamics from the
exact values of the distribution, except that they must be greater than zero. 
It is worth noting that
effective dynamics is not always non linear, as 
shown in~\sref{subsec:nonmarko}. Interestingly, this example is non linear and
non-Markovian; thus, closed (and trivially Markovian) microscopic dynamics leads
to non-Markovianity.

The proposed framework has both fundamental and practical applications. On the
fundamental side, it contributes to understanding how effective nonlinearity
can emerge from the combination of linear Schrödinger dynamics and addressing
errors. This perspective may offer insights into the quantum-to-classical
transition at the level of dynamical evolution. However, further investigations
are needed to explore this connection in greater depth. On the practical side,
when analyzing the dynamics of quantum many-body systems—such as spin
lattices~\cite{Eisert2015} or IBM quantum processors with various hardware
geometries~\cite{Kandala2017}—the framework provides a principled basis for identifying
and quantifying a source of decoherence. Notably, in systems with many
particles, the cumulative effect of addressing errors can become significant
even when individual error probabilities remain small, due to the exponential
sensitivity~\cite{FuzzyMeasurements}.

Finally, several questions arise from this work. 
For example, under what conditions does the
emergent non linearity depend only on the distribution of the coarse graining?
Inspired by the Ising spin chain, is it possible to find non linear dynamics
that is independent of the initial state?

\section*{Acknowledgments} It is with great pleasure that we thank Fernando de Melo, Raul Vallejo, and Hamed Mohammady
for their useful comments and suggestions. We acknowledge project VEGA No. 2/0183/21 (DESCOM).
This work was supported by UNAM-PAPIIT Grant No. IG101324.
\section*{Data availability}
The data that supports the findings of this article are openly available~\cite{data}.
\bibliographystyle{unsrt} 
 
\appendix 
\section{Fuzzy operators}\label{ap:FuzzyOperators}
In this appendix we show the relation between the expected values
of a tomographically complete set of observables on the effective
Hilbert space $\{\varsigma^{\alpha}\}_{\alpha}$ and the expected 
values of what we call fuzzy operators.
Recalling that the effective and the microscopic states are
related through $\rho_{\ef}=\mcC(\varrho)$, we can write
\begin{align}
    \tr(\varsigma^{\alpha}\rho_{\ef})&=\tr[\varsigma^{\alpha}\mcC(\varrho)]\nonumber\\
    &=\tr[\varsigma^{\alpha}\tr_{\overline{1}}\qty(\sum_{k=1}^{n}p_{k}P_{1,k}\varrho P_{1,k})]\nonumber\\
    &=\tr[\varsigma^{\alpha}_{1}\sum_{k=1}^{n}p_{k}P_{1,k}\varrho P_{1,k}]\\
    &=\sum_{k=1}^{n}p_{k}\tr(\varsigma^{\alpha}_{1}P_{1,k}\varrho P_{1,k})\\
    &=\sum_{k=1}^{n}p_{k}\tr(P_{1,k}\varsigma^{\alpha}_{1}P_{1,k}\varrho )\\
    &=\tr[\left(\sum_{k=1}^{n}p_{k} P_{1,k} \varsigma^{\alpha}_{1}P_{1,k}\right)\varrho ]\\
    &=\tr[\left(\sum_{k=1}^{n}p_{k}\varsigma^{\alpha}_{k}\right)\varrho ],
\end{align}
where we have used the fact that swaps $P_{j,k}$ are Hermitian and the cyclic property of the trace. Additionally, $\varsigma^{\alpha}_k$ is the observable $\varsigma^\alpha$ applied to the $k$th particle.
Finally, we can identify the operators $G^\alpha$ as defined in~\eqref{eq:GOperators} where the following holds:
\begin{equation}
    \tr(\varsigma^{\alpha}\rho_{\ef})=\tr(G^{\alpha}\varrho).
\end{equation}
\section{Elliptic \textsc{cnot} path} \label{ap:EllipticCNOT}
Here we show that the interaction term of the \textsc{cnot} Hamiltonian, 
$\sigma^{z}\sigma^{x}$, results in an elliptical path for the 
Bloch vector of the effective state. First, it is easy to show 
\cite{EllipticalOrbits} that if we assume that the initial 
microscopic state is a product state, then the reduced Bloch 
vectors evolve according to
\begin{equation}
    \begin{aligned}
        r_{1}^{x}(t)=&r_{1}^{x}(0)\cos(t)+r_{1}^{y}(0)r_{2}^{x}(0)\sin(t),\\
        r_{1}^{y}(t)=&r_{1}^{y}(0)\cos(t)-r_{1}^{x}(0)r_{2}^{x}(0)\sin(t),\\
        r_{1}^{z}(t)=&r_{1}^{z}(0)
    \end{aligned}
\end{equation}
and
\begin{equation}
    \begin{aligned}
    r_{2}^{x}(t)=&r_{2}^{x}(0),\\
    r_{2}^{y}(t)=&r_{2}^{y}(0)\cos(t)-r_{1}^{z}(0)r_{2}^{z}(0)\sin(t),\\
    r_{2}^{z}(t)=&r_{2}^{z}(0)\cos(t)+r_{1}^{z}(0)r_{2}^{y}(0)\sin(t),
\end{aligned}
\end{equation}
which correspond to ellipses confined to the $x$-$y$ and $y$-$z$ planes.
The Bloch vector of the effective state is of course the sum of
the Bloch vectors of the two qubits, and thus the effective Bloch
vector evolves according to
\begin{equation}
    \vec{r}_{\ef}=p_{1}\vec{r}_{1}+p_{2}\vec{r}_{2}.
\end{equation}
As it turns out, this is also an ellipse, as it can be written as
\begin{equation}
    \vec{r}_{\ef}=\vec{u}\sin(t)+\vec{v}\cos(t)+\vec{c}.
\end{equation}
Here,
\begin{align}
    \vec{u} = 
    \begin{bmatrix} -b_1p_1\sin(\theta_1) \\ 
        p_1 [a_1 \sin(\theta_1) + b_1 \cos(\theta_1)]\\ 
        p_2 [a_2 \sin(\theta_2) + b_2 \cos(\theta_2)]
    \end{bmatrix}, 
\end{align}
$\vec{c} = \left[ p_2 x_{2}(0) \quad 0 \quad p_1 z_{1}(0) \right]^T$,
and
\begin{equation}
    \vec{v} = 
    \begin{bmatrix} 
        a_1p_1\cos(\theta_1) \\ 
        p_2[a_2 \cos(\theta_2) - b_2 \sin(\theta_2)] \\ 
        0 
    \end{bmatrix}.
\end{equation}
The parameters $a_{j}$, $b_{j}$, and $\theta_{j}$ correspond
to the ellipse parameters of the path followed by each reduced
system and are related to their initial Bloch vectors through
\begin{align}
    a_{1}\cos(\theta_{1})=r_{1}^{x}(0), & & a_{1}\sin(\theta_{1})=-r_{1}^{y}(0),\nonumber\\
    b_{1}\cos(\theta_{1})=-r_{1}^{x}(0)r_{2}^{x}(0), & & b_{1}\sin(\theta_{1})=r_{1}^{y}(0)r_{2}^{x}(0)\nonumber
\end{align}
and
\begin{align}
    a_{2}\cos(\theta_{2})=r_{2}^{y}(0), & & a_{2}\sin(\theta_{2})=-r_{2}^{z}(0),\nonumber\\
    b_{2}\cos(\theta_{2})=r_{1}^{z}(0)r_{2}^{y}(0), & & b_{2}\sin(\theta_{2})=-r_{1}^{z}(0)r_{2}^{z}(0)\nonumber.
\end{align}
\section{Ising Model without magnetic field} \label{app:ising}
As mentioned in the main text, to compute the coarse-grained state of the spin chain after the unitary evolution of the microscopic description with $g/J=0$, it is enough to compute the reduced density matrix for any spin. To do it, first observe that due to the translation symmetry and the specific form of the initial state $\ket{\psi}^{\otimes N}$, the reduced evolution for each spin is identical. The density matrix of the initial pure state of the spin tagged with $N-1$, $\ket{\psi}=\cos\left( \theta/2 \right) \ket{0}+e^{i \phi} \sin \left( \theta/2\right) \ket{1}$, in the computational basis is
$$\rho_{N-1}(0)=\left(
\begin{array}{cc}
 \cos ^2\left(\frac{\theta }{2}\right) & \frac{1}{2} e^{-i \phi } \sin (\theta ) \\
 \frac{1}{2} e^{i \phi } \sin (\theta ) & \sin ^2\left(\frac{\theta }{2}\right) \\
\end{array}
\right).$$
and the total density matrix of the microscopic description, also in the computational basis, is 
\begin{multline}
\left(\ketbra{\psi}\right)^{\otimes N}=\\
\sum_{\vec k, \vec l} C^N_kC^N_l e^{i\left(\phi+\pi/2\right)(l-k)}c^{k+l}_\theta 
s_\theta^{2N-k-l}  \ketbra{\vec k}{\vec l}.
\label{eq:total_state_spin_expanded}
\end{multline}
with $C^N_k$ and $C^N_l$ the binomial coefficients where $k$ and $l$ are the numbers of $0$'s in $\vec k$ and $\vec l$, respectively. We also use the abbreviations $c_\theta:=\cos(\theta/2)$ and $s_\theta:=\sin(\theta/2)$.
To derive its evolution it is enough to compute the components $\left[\rho_{N-1}(t)\right]_{00}$ and $\left[\rho_{N-1}(t)\right]_{01}$. Now, observe the following: When tracing out all spins except $N-1$, only operators with the form $\ketbra{0\vec i}$ contribute to $\ketbra{0}$, where $\vec i$ is a vector indicating the rest of the $0$'s and $1$'s of the computational basis. Since $\ket{0\vec i}$ is an eigenstate of the Hamiltonian, the component $\left[\rho_{N-1}(t)\right]_{00}=\left[\rho_{N-1}(0)\right]_{00}$ remains invariant. 

For $\left[\rho_{N-1}(t)\right]_{01}$ only operators with the form $\ketbra{0\vec i}{1\vec i}$ contribute to the partial trace. Thus, let us investigate their evolution. Observe that both $\ket{0\vec i}$ and $\ket{1\vec i}$ are eigenvectors of the evolution operator of the spin chain. Therefore, assume that 
$$U(t)\ket{0\vec i}=e^{i t E_{0\vec i}}\ket{0\vec i},$$ 
where $E_{0\vec i}$ is the eigenenergy of $\ket{0\vec i}$. Now, we want to find the relative phase with $\ket{1\vec i}$. To do this it is enough to observe what happens with the nearest neighbors of $N-1$ (spins $0$ and $N-2$). So we need to investigate the changes in just four configurations: $0\mathbf{0}0$. Starting with $000$, where the spin at the center is $N-1$ and others are its neighbors, changing it to $010$ ``removes'' two $+J$ terms in the eigenenergy expression of state $\ket{00\vec i'0}$ in favor of two $-J$ terms; thus 
$$U(t)\ket{10\vec i'0}=e^{ i t \left(E_{0\vec i}-4J\right)}\ket{10\vec i'0},$$
where the left most spin is the one tagged with $N-1$ and the other spins indicated explicitly are its neighbors. The vector $\vec i'$ contains the rest of the spins in the computational basis.
Putting both kets together we have 
$$U(t)\ketbra{00\vec i'0}{10\vec i' 0}U(-t)=e^{i t 4J}\ketbra{00\vec i'0}{10\vec i' 0}.$$
To compute how operators $\ketbra{00\vec i'}{10\vec i' 0}$ contribute to the partial trace, observe that they appear in the initial density matrix weighted with 
$$e^{-i \phi}\cos^5{\left(\frac{\theta}{2}\right)}\sin{\left(\frac{\theta}{2}\right)} \cos^{2k}{\left(\frac{\theta}{2}\right)}\sin^{2(N-3-k)}{\left(\frac{\theta}{2}\right)},$$
where $k$ is the number of $0s$ in $\vec i'$ [see~\eref{eq:total_state_spin_expanded}]. Moreover, after tracing out all spins except the one tagged with $N-1$, there are exactly $C^{N-3}_k=(N-3)!/k! (N-3-k)!$ such factors for each $k$. Therefore, the total contribution to the operator $\ketbra{0}{1}$ from the family of operators with the form $\ketbra{00\vec i'0}{10\vec i' 0}$ is
\begin{align}
&e^{-i \phi}\cos^5{\left(\frac{\theta}{2}\right)}\sin{\left(\frac{\theta}{2}\right)}\nonumber\\ &\times \sum_{k=0}^{N-3} C^{N-3}_k \cos^{2k}{\left(\frac{\theta}{2}\right)}\sin^{2(N-3-k)}{\left(\frac{\theta}{2}\right)}\nonumber \\
=&e^{-i \phi}\cos^5{\left(\frac{\theta}{2}\right)}\sin{\left(\frac{\theta}{2}\right)}\nonumber\\ &\times \left(\cos^2\left(\frac{\theta}{2}\right)+\sin^2\left(\frac{\theta}{2}\right) \right)^{N-3}\nonumber\\
= &e^{-i \phi}\cos^5{\left(\frac{\theta}{2}\right)}\sin{\left(\frac{\theta}{2}\right)}.
\end{align} 
This result, together with the rest of the cases, is summarized in the Table~\ref{tab:table}.
\begin{table}
\centering
{\renewcommand{\arraystretch}{1.5}
\begin{tabular}{ccc}
\hline 
Operator family & Contribution & Phase \\ 
\hline 
$\ketbra{\mathbf{0}0\vec i' 0}{\mathbf{1}0\vec i' 0}$ & $e^{-i\phi}c_\theta^5 s_\theta$ & $e^{it4J}$ \\ 
$\ketbra{\mathbf{0}0\vec i' 1}{\mathbf{1}0\vec i' 1}$ & $e^{-i \phi}c_\theta^3 s_\theta^3$ & 1 \\ 
$\ketbra{\mathbf{0}1\vec i' 0}{\mathbf{1}1\vec i' 0}$ & $e^{-i \phi}c_\theta^3 s_\theta^3$ & 1 \\ 
$\ketbra{\mathbf{0}1\vec i' 1}{\mathbf{1}1\vec i' 1}$ & $e^{-i\phi}c_\theta s_\theta^5$ & $e^{-it 4J}$ \\ 
\end{tabular} 
}
\caption{Summary of phases gained by each operator type (parametrized by $\vec i'$) during microscopic evolution; the $(N-1)$th spin is in bold. The contribution is the sum of all factors that remain once all operators of each family are partially traced (see the main text for details). We use the abbreviations $c_\theta:=\cos(\theta/2)$ and $s_\theta:=\sin(\theta/2)$. \label{tab:table}}
\end{table}
Adding up all contributions to the operator $\ketbra{0}{1}$, we have (using the notation of the Table~\ref{tab:table})
\begin{align}
&\left[\rho_{N-1}(t)\right]_{01}=\nonumber\\
&2\left( e^{i t4J}c_\theta^5 +2c_\theta^3 s_\theta^2+e^{-i t 4J}c_\theta s_\theta^4  \right) \frac{e^{-i \phi}s_\theta}{2} \nonumber\\
&=\gamma(\theta,t)\left[\rho_{N-1}(0)\right]_{01},
\end{align}
where we have identified $e^{-i \phi}s_\theta/2=\left[\rho_{N-1}(0)\right]_{01}$. The factor $\gamma(\theta,t)$ can be further simplified [see~\eref{eq:gamma_ising}]. Notice that the result is independent of $N$.
\section{Neglectability of terms in the power series of effective all-to-all interaction}\label{ap:small_commutator} 

In~\sref{sec:SpinChains}, chains evolving due
to a non uniform external magnetic field in the $z$ direction
with an all-to-all Ising interaction parallel to the field were
considered. By iteratively integrating the Liouville–von Neumann equation,
a Dyson series, given by~\eref{eq:Dyson-like}, was obtained.
In this series, terms proportional to $\mcC\{[H_\text{int},\varrho_{\max}(0)]\}$
arise, which are discarded. In this appendix we show that they are in fact negligible.

By virtue of~\eref{eq:MaxEntAss}, we know that $\varrho_{\max}(0)$
is of the form $\Motimes \rho_k$ and so
\begin{equation}
    \begin{aligned}
        \mcC\{[H_\text{int},  \varrho_{\max}(0)]\}& \\
        = \sum_{k=1}^{N}p_k[\rho_k,\sigma^z]&\prod_{j\neq k}\tr(\rho_k\sigma^z) \\
        = \prod_{j=1}^{N}\tr(\rho_j\sigma^z)& \sum_{k=1}^{N}\frac{p_k}{\tr(\rho_k\sigma^z)}[\rho_k,\sigma^z],
    \end{aligned}
\end{equation}
meaning that
\begin{equation}\label{eq:commutatorpropto}
    \mcC\{[H_\text{int},\varrho_{\max}(0)]\}\propto\prod_{j=1}^{N}\tr(\rho_j\sigma^z).
\end{equation}
The product on the right-hand side of~\eref{eq:commutatorpropto} is a product of
the $z$ components of each subsystem of $\varrho_{\max}(0)$, which we know to
be
\begin{equation}
    \tr(\rho_j\sigma^z)=\frac{r^{z}_{\text{eff} }}{r_\text{eff}}\tanh(\lambda p_j),
\end{equation}
where $r^{z}_{\text{eff}}/r_\text{eff}<1$. Now, as $N$ grows, 
$\tanh(\lambda p_j) \approx \lambda p_j$, and if we take $p_j$ of $\mcO(1/N)$, 
then the product
\begin{equation}
    \prod_{j=1}^{N}\tr(\rho_j\sigma^z) 
= \mcO \left( \frac{1}{N^N} \right),
\end{equation}
which implies that
\begin{equation}\label{eq:commutatorproptofinal}
    \mcC\{[H_\text{int},\varrho_{\max}(0)]\}
= \mcO \left( \frac{1}{N^N} \right),
\end{equation}
which decays exponentially as $N$ grows. We conclude that once the coarse-graining 
map is applied, the odd terms in \eqref{eq:CommutatorCases} can be approximated by
$0$. 

\end{document}